\documentclass[11pt]{article}

\usepackage{fontspec}
\defaultfontfeatures{Mapping=tex-text}
\usepackage{xunicode}
\usepackage{xltxtra}
\usepackage[a4paper, margin=2cm]{geometry}

\usepackage[english]{babel}

\usepackage{amsmath, amssymb, bm}
\usepackage{slashed}

\usepackage{graphicx}
\usepackage{float}
\usepackage{subcaption}
\usepackage{booktabs}
\usepackage{siunitx}
\usepackage{multirow}

\usepackage{titlesec}
\titleformat{\section}
{\normalfont\Large\bfseries\centering}{\Roman{section}.}{1em}{}

\usepackage{cite}
\usepackage{indentfirst}
\usepackage{appendix}

\begin{document}

% 手动创建标题，避免 \title 命令的问题
\begin{center}
{\LARGE \textbf{Order-$v^4$ corrections to  heavy quark fragmentation to S-wave heavy quarkonium}} \\[12pt]
%{\large Your Name} \\[8pt]
%{\small Your Institution} \\[8pt]
  \large % 作者字号略大，符合标题页排版习惯
  Sai Cui$^{(a)}$, Yi-Jie Li$^{(a)}$, Guang-Zhi Xu$^{(a),*}$ and Kui-Yong Liu$^{(b,a),\dagger}$
  
  \vspace{6pt} % 作者与单位间的间距（替代之前的\\[6pt]）
  
  \normalsize % 单位字号恢复常规
  $^{(a)}$School of Physics, Liaoning University, Shenyang 110036, China\\
  $^{(b)}$School of Physics and Electronic Technology, Liaoning Normal University, Dalian 116029, China
  
  \vspace{6pt} % 单位与通讯作者间的间距
  
  $*$Corresponding author. Email: xuguangzhi@lnu.edu.cn\\
  $\dagger$Corresponding author. Email: liukuiyong@lnu.edu.cn

{\small \today}
\end{center}

\begin{abstract}
	
	Within the framework of nonrelativistic quantum chromodynamics (NRQCD) factorization, we compute the
	$\mathcal{O}(v^{4})$ relativistic corrections to the fragmentation of a heavy quark into the color-singlet
	$^{1}S_{0}^{[1]}$ and $^{3}S_{1}^{[1]}$ quarkonium states.
	Using the Collins--Soper definition of the fragmentation function, we reproduce the known
	$\mathcal{O}(v^{2})$ results.
	We find that the $\mathcal{O}(v^{4})$ correction gives a positive contribution relative to the leading order result
	over a wide range of the light-cone momentum fraction $z$, while its magnitude remains much smaller than that of the
	$\mathcal{O}(v^{2})$ correction.
	This behavior indicates a good convergence of the NRQCD relativistic expansion in this process.
We further extend the calculation to the fragmentation functions in the unequal‑mass case at $\mathcal{O}(v^{4})$ and obtain the corresponding analytical expressions.

	\vspace{0.5cm}
\end{abstract}

\section{Introduction}

Heavy quarkonium has long served as a crucial probe for testing strong interactions in high-energy collision experiments. Existing experimental data reveal persistent discrepancies between measured cross sections and polarization observables in the high transverse-momentum region and current theoretical predictions. This situation underscores the necessity of higher-precision theoretical studies of quarkonium production mechanisms, motivating extensive investigations of higher-order QCD and relativistic effects across different production channels~\cite{ Brambilla:2010cs, Kang:2011mg, Bodwin:2008vp, Gong:2008ft, Ma:2008gq, Zhang:2009ym, He:2009uf, Guo:2011tz, Li:2012rn, Xu:2012am, Li:2013csa, Xu:2014zra }.

The Nonrelativistic QCD (NRQCD) factorization framework provides a systematic approach by separating perturbatively calculable short-distance coefficients (SDCs) from non-perturbative long-distance matrix elements (LDMEs) that describe bound-state formation~\cite{BBL1995}, based on the effective field theory formulation of NRQCD~\cite{CaswellLepage1986}. In the high transverse-momentum region, the cross section can be expressed as a convolution of a hard-scattering cross section and fragmentation functions~\cite{Collins:1981uw,Collins1989}. Among these, the gauge-invariant fragmentation function defined by Collins and Soper forms the foundation for describing heavy quarkonium production via the fragmentation mechanism~\cite{Collins:1981uw}.

Within this framework, the fragmentation of partons (quarks and gluons) is considered the dominant production mechanism at high momentum. Extensive earlier work established the leading-order (LO) fragmentation functions for quarkonia with different angular momenta~\cite{Ma:1994zt,Ma:1995ci,Braaten:1993mp,Ma:1995vi,Yuan:1994hn,Braaten:1993rw,Braaten:1994kd,Cheung:1995ir,Bodwin:2014bia,Cho:1994qp,Qiao:1997wb}, laying the foundation for subsequent higher-order studies. As experimental precision improves, QCD corrections to fragmentation functions have become a major focus. Existing studies have systematically investigated next-to-leading-order (NLO) QCD corrections for heavy quark and gluon fragmentation into $S$-wave and $P$-wave states~\cite{Zhang:2018mlo,Zheng:2019dfk,Sepahvand:2017gup,Zheng:2021ylc,Feng:2021uct,Feng:2018ulg}, including crucial two-loop contributions~\cite{Zhang:2020atv} that have significantly enhanced the reliability of theoretical predictions.

In addition to QCD radiative corrections, relativistic effects in fragmentation channels are equally significant. Current research has demonstrated that the $\mathcal{O}(v^2)$ relativistic corrections to parton fragmentation functions provide notable contributions to the LO results~\cite{Sang:2009zz,Gao:2016ihc,Bodwin:2003wh}, raising questions about the convergence of the NRQCD relativistic expansion. To assess the validity of this expansion, several studies have extended calculations to $\mathcal{O}(v^4)$. These include complete $\mathcal{O}(v^4)$ corrections for $S$-wave decay processes~\cite{Bodwin:2002cfe}, $\mathcal{O}(v^4)$ contributions to $\Upsilon$ decays~\cite{Sang:2012yh}, and the $\mathcal{O}(v^4)$ structure for gluon fragmentation into ${}^{3}S_{1}$ states~\cite{Bodwin:2012xc}. Notably, Ref.~\cite{Bodwin:2012xc} revealed, for the first time at $\mathcal{O}(v^4)$, an infrared-divergent structure, suggesting that higher-order relativistic effects may have special significance within the factorization framework. Furthermore, $\mathcal{O}(v^4)$ corrections have been shown to yield significant effects in processes such as Higgs decay into $J/\psi + \gamma$~\cite{Brambilla:2019fmu}.

Despite extensive studies on $\mathcal{O}(v^{4})$ corrections for various processes, a systematic calculation of the complete SDCs at this order for heavy quark fragmentation into $S$-wave quarkonia is still lacking. To fill this gap, we perform, for the first time, a calculation of the $\mathcal{O}(v^{4})$ relativistic corrections for this process and obtain the corresponding SDCs by matching with the standardly normalized NRQCD matrix elements. These results provide the necessary theoretical input to improve the precision of predictions in the high-momentum region and to further test the NRQCD factorization framework.

The remainder of this paper is organized as follows. In Section~\ref{CS}, we review the Collins--Soper definition of the fragmentation function. In Section~\ref{NRQCD}, we introduce the NRQCD factorization framework. In Section~\ref{axial}, we detail the calculation of the fragmentation function for a free quark-antiquark pair in the axial gauge. In Section~\ref{Matching}, we present the matching of the fragmentation function onto NRQCD operators. In Section~\ref{Num}, we show our numerical results and analyze the $\mathcal{O}(v^{4})$ corrections. In Section~\ref{summary}, we provide a summary. In Appendix~A, we present the order-by-order expansion of the squared amplitude. In Appendix~B, we compute the SDCs for the unequal-mass case up to $\mathcal{O}(v^{4})$.

\section{Collins-Soper definition of fragmentation}
\label{CS}

This section introduces the gauge-invariant definition of the fragmentation function given by Collins and Soper \cite{Collins:1981uw}. Based on this definition, Sang et al. have calculated the SDCs for heavy quark fragmentation into equal-mass and unequal-mass $S$-wave quarkonia up to $\mathcal{O}(v^2)$ corrections \cite{Sang:2009zz}. 

We introduce light-cone coordinates for calculational convenience.
In light-cone coordinates, a four-vector $p^\mu$ is expressed as
\begin{align}
	 &p= (p^+, p^-, \mathbf{p_T}) , \label{eq:p_def} \\
 &	p^+ = (p^0 + p^3)/\sqrt{2}, \label{eq:p_plus} \\
	 &p^- = (p^0 - p^3)/\sqrt{2}. \label{eq:p_minus}
\end{align}
The scalar product of two four-vectors $p$ and $q$ is then
\begin{equation}
p \cdot q = p^+ q^- + p^- q^+ -\mathbf{p_T} \cdot\mathbf{q_T} .
\end{equation}
To describe the light-cone propagation in the fragmentation function, we introduce an auxiliary vector in the “minus” light-cone direction
\begin{equation}
	\hat{n}^{\mu} = (0,\,1,\,\mathbf{0_T})
\end{equation}
 with $\hat{n}^2 = 0$.

Adopting the gauge-invariant definition by Collins–Soper, the fragmentation function for a heavy quark into a hadron $H$ is written as \cite{Collins:1981uw}
\begin{equation}
	\begin{aligned}
		D_{H/Q}(z)
		&= \frac{z^{\,d-3}}{4\pi}
		\int dx^{-}\, e^{- i P^{+} x^{-}/z}\,
		\frac{1}{3}\mathrm{Tr}_{\text{color}}
		\,\frac{1}{2}\mathrm{Tr}_{\text{Dirac}}
		\\[4pt]
		&\times
		\Biggl\{
		\hat{n}\!\cdot\!\gamma\;
		\langle 0|
		Q(0)\,
		\bar {\text{P}} \exp\!\left[
		-i g_{s}\!\int_{0}^{\infty}\! d\lambda\;
		\hat{n}\!\cdot\! A^{T}(\lambda \hat{n}^\mu)
		\right]
		\\[4pt]
		&\times
		a_{H}^{\dagger}(P^{+},\mathbf{0_{T}})\,
		a_{H}(P^{+},\mathbf{0_{T}})\,
	\text{P}\exp\!\left[
		i g_{s}\!\int_{x^{-}}^{\infty}\! d\lambda\;
		\hat{n}\!\cdot\! A^{T}(\lambda \hat{n}^\mu)
		\right]
		\\[4pt]
		&\times
		\bar Q(0, x^{-}, \mathbf{0_{T}})
		|0\rangle
		\Biggr\}.
		\qquad 
	\end{aligned}
\end{equation}
Here, $z = P^+ / k^+$ is the fraction of the heavy quark's  $+$  component of momentum that is carried by the hadron $H$ with momentum $P$. The parameter $d = 4 - 2\epsilon$ denotes the space-time dimension in dimensional regularization. Since our calculation does not involve divergent terms, we set $d = 4$ in subsequent computations. The term \(\text{P}\exp\{\cdots\} \) ensures the gauge invariance of the operator. $a_{H}^{\dagger}$ represents the creation operator for the hadron $H$ and $A_{\mu}$ is the gluon field.

This definition serves as the foundation for subsequent NRQCD matching calculations and provides a directly applicable gauge-invariant framework for Sang et al.'s relativistic expansion of the short-distance coefficients for $S$-wave quarkonia \cite{Sang:2009zz}.

\section{NRQCD factorization }
\label{NRQCD}

Within the framework of NRQCD factorization, the fragmentation function for a heavy quark $Q$ fragmenting into a hadron $H$ can be expressed as \cite{Collins:1981uw, Collins1989}
\begin{equation}
	\label{ex}
	\begin{aligned}
		D_{H/Q}(z) = \sum_{n} \Bigl[ & F^{n}(z) \langle 0| \mathcal{O}^{H}(n) |0 \rangle + G^{n}(z) \langle 0| \mathcal{P}^{H}(n) |0 \rangle \\
		& + H_1^{n}(z) \langle 0| \mathcal{Q}_1^{H}(n) |0 \rangle  + H_2^{n}(z) \langle 0| \mathcal{Q}_{2}^{H}(n) |0 \rangle \Bigr] + O(v^{6}).
	\end{aligned}
\end{equation}
where \(\langle 0| 	\mathcal O^{H}(n) |0 \rangle\), \(\langle 0|	\mathcal P^{H}(n) |0 \rangle\), and \(\langle 0|	\mathcal Q_i^{H}(n) |0 \rangle(i=1,2,3)\) are the NRQCD LDMEs, \(F^{n}(z)\), \(G^{n}(z)\), and \(H_i^{n}(z)(i=1,2,3)\) are the corresponding SDCs, and \(n\) labels the quantum numbers of the intermediate \(Q\bar{Q}\) pair. It is the quantity \( H_1^{n}(z)+ H_2^{n}(z)\) that we compute in this paper.

For a hadron \(H\) with quantum numbers \( ^1 S_0 \), the relevant operators are defined as
\begin{align}
	\mathcal{O}^{H}(^1 S_0) &= \frac{1}{N_c} \chi^{\dagger} \psi \sum_X |H + X\rangle \langle H + X| \psi^{\dagger} \chi, \\
	\mathcal{P}^{H}(^1 S_0) &= \frac{1}{2N_c} \bigg[ \chi^{\dagger} \bigg( \frac{i \overleftrightarrow{\mathbf{D}}}{2} \bigg)^2 \psi \sum_X |H + X\rangle \langle H + X| \psi^{\dagger} \chi + \mathrm{H.c.} \bigg],
\end{align}
\begin{align}
	\mathcal{Q}_{1}^H (^1S_0) &= \frac{1}{N_c}\chi^\dagger \left( -\frac{i}{2} \overleftrightarrow{\mathbf{D}} \right)^2 \psi \sum_X |H + X\rangle \langle H + X| \psi^\dagger \left( -\frac{i}{2} \overleftrightarrow{\mathbf{D}} \right)^2 \chi, \\
	\mathcal{Q}_{2}^H (^1S_0) &= \frac{1}{2N_c} \left[ \chi^\dagger \left( -\frac{i}{2} \overleftrightarrow{\mathbf{D}} \right)^4 \psi \sum_X |H + X\rangle \langle H + X| \psi^\dagger \chi + \mathrm{H.c.} \right], \\
	\mathcal{Q}_{3}^H (^1S_0) &= \frac{1}{2N_c} \left[ \chi^\dagger \psi \sum_X |H + X\rangle \langle H + X| \psi^\dagger (\overleftrightarrow{\mathbf{D}} \cdot g_s \mathbf{E} + g_s \mathbf{E} \cdot \overleftrightarrow{\mathbf{D}}) \chi \right. \\
	&\quad \left. - \chi^\dagger (\overleftrightarrow{\mathbf{D}} \cdot g_s \mathbf{E} + g_s \mathbf{E} \cdot \overleftrightarrow{\mathbf{D}}) \psi \sum_X |H + X\rangle \langle H + X| \psi^\dagger \chi \right].
\end{align}
For a hadron \( H \) with quantum numbers \( ^3 S_1 \), the relevant operators are defined as
\begin{align}
	\mathcal{O}^{H}(^3 S_1) &= \frac{1}{N_c} \chi^{\dagger} \sigma^{i} \psi \sum_X |H + X\rangle \langle H + X| \psi^{\dagger} \sigma^{i} \chi, \\
	\mathcal{P}^{H}(^3 S_1) &= \frac{1}{2N_c} \left[ \chi^{\dagger} \sigma^{i} \left( \frac{i \overleftrightarrow{\mathbf{D}}}{2} \right)^2 \psi \sum_X |H + X\rangle \langle H + X| \psi^{\dagger} \sigma^{i} \chi + \mathrm{H.c.} \right],
\end{align}
\begin{align}
	\mathcal{Q}_{1}^H(^3S_1) &= \frac{1}{N_c} \chi^\dagger\sigma^i\left(-\frac{i}{2}\overleftrightarrow{\mathbf{D}}\right)^2\psi \sum_X |H + X\rangle \langle H + X| \psi^\dagger\sigma^i\left(-\frac{i}{2}\overleftrightarrow{\mathbf{D}}\right)^2\chi, \\
	\mathcal{Q}_{2}^H(^3S_1) &= \frac{1}{2N_c} \left[ \chi^\dagger\sigma^i\left(-\frac{i}{2}\overleftrightarrow{\mathbf{D}}\right)^4\psi \sum_X |H + X\rangle \langle H + X| \psi^\dagger\sigma^i\chi + \mathrm{H.c.} \right], \\
	\mathcal{Q}_{3}^H(^3S_1) &= \frac{1}{2N_c} \left[ \chi^\dagger\sigma^i\psi \sum_X |H + X\rangle \langle H + X| \psi^\dagger\sigma^i(\overleftrightarrow{\mathbf{D}} \cdot g_s \mathbf{E} + g_s \mathbf{E} \cdot \overleftrightarrow{\mathbf{D}})\chi \right. \\
	&\quad \left. - \chi^\dagger\sigma^i(\overleftrightarrow{\mathbf{D}} \cdot g_s \mathbf{E} + g_s \mathbf{E} \cdot \overleftrightarrow{\mathbf{D}})\psi \sum_X |H + X\rangle \langle H + X| \psi^\dagger\sigma^i\chi \right].
\end{align}

$H$ denotes the heavy-quarkonium hadronic state, 
$X$ represents the accompanying produced particles, and the summation 
$\sum_X |H+X\rangle\langle H+X|$ corresponds to the projection operator onto the final-state subspace containing $H$. 
$N_c = 3$ is the number of colors. 
$\psi$ and $\chi$ are the two-component Pauli spinor fields for the heavy quark and the antiquark, respectively. 
$\sigma^i$ are the Pauli matrices, and $\mathbf{D}$ denotes the spatial part of the covariant derivative. 
The combination $\overleftrightarrow{\mathbf{D}} \equiv \overrightarrow{\mathbf{D}} - \overleftarrow{\mathbf{D}}$ 
serves as the basic building block for the expansion in relativistic corrections. 
$g_s$ is the strong coupling constant, and $\mathbf{E}$ is the chromoelectric field operator. 
The operators defined above correspond to the leading contribution $\mathcal{O}^{H}$, 
the relativistic correction at order $v^{2}$ denoted by $\mathcal{P}^{H}$, and the 
order $v^{4}$ contributions represented by $\mathcal{Q}_{1}^{H}$, $\mathcal{Q}_{2}^{H}$, 
and $\mathcal{Q}_{3}^{H}$.  
The operator $\mathcal{Q}_{3}^{H}$ encodes the interaction between the heavy-quark pair 
and the chromoelectric field.

A detailed analysis in Ref.~\cite{Bodwin:2002cfe} demonstrates that the matrix element of 
$\mathcal{Q}_{3}^{H}$ is not independent at relative order $v^{4}$.  
Its contribution can be rewritten in terms of the matrix elements associated with 
$\mathcal{Q}_{1}^{H}$ and $\mathcal{Q}_{2}^{H}$, meaning that the removal of 
$\langle \mathcal{Q}_{3}^{H} \rangle$ does not affect the completeness of the NRQCD operator basis at this accuracy.
In potential-model estimates and within the vacuum-saturation approximation, the two 
remaining order $v^{4}$ operators satisfy the approximate relation
\begin{equation}
	\langle 0|\mathcal{Q}_{1}^{H}(n)|0\rangle 
	\approx 
	\langle 0|\mathcal{Q}_{2}^{H}(n)|0\rangle ,
	\label{eq:Q1Q2-rel}
\end{equation}
so that their SDCs enter only through a single effective 
combination
\begin{equation}
	H_1^{n}(z)\,\langle 0|\mathcal{Q}_{1}^{H}(n)|0\rangle
	+
	H_2^{n}(z)\,\langle 0|\mathcal{Q}_{2}^{H}(n)|0\rangle
	\approx
	\bigl[ H_1^{n}(z) + H_2^{n}(z) \bigr]\,
	\langle 0|\mathcal{Q}_{1}^{H}(n)|0\rangle .
	\label{eq:H1H2-combine}
\end{equation}
Therefore, at order $v^{4}$, the analysis requires only the combined short-distance 
coefficient $H_{1}(z)+H_{2}(z)$, and the independent computation of the coefficient 
associated with $\mathcal{Q}_{3}^{H}$ is unnecessary.

\section{ Calculating the fragmentation functions in the axial gauge}
\label{axial}
In this section, we compute the fragmentation function for a heavy quark fragmenting into $S$-wave hadronic states up to $\mathcal{O}(v^4)$ relativistic corrections in the axial gauge. The momenta of the quark $Q$ and antiquark $\bar{Q}$ can be expressed as
\begin{equation}
	p_1 = \frac{P}{2} + q, \qquad p_2 = \frac{P}{2} - q,
\end{equation}
where $P$ is the hadron momentum and $q$ the relative momentum. In the rest frame of the $Q\bar{Q}$ pair, the components are chosen as
\begin{equation}
	P = (2E, \mathbf{0}), \qquad q = (0, \mathbf{q}),
\end{equation}
so that the three‑momentum of the heavy quark is $\mathbf{q}$ and the invariant mass of the $Q\bar{Q}$ state satisfies
\begin{equation}
	P^{2} = M^{2} = 4E^{2}, \qquad E = \sqrt{m^{2} - q^{2}} = \sqrt{m^{2} + \mathbf{q}^{2}},
\end{equation}
with $m$ being the heavy‑quark mass. In the axial gauge $\hat{n}\!\cdot\!A = 0$, the Wilson line in the Collins--Soper definition becomes trivial.
Consequently, all eikonal-gluon contributions are absent, and only diagrams with a single gluon exchanged
between the fragmenting parton and the heavy-quark pair contribute to the fragmentation function.
According to Ref.~\cite{Bodwin:2002cfe}, we introduce the projection operators for spin and color.
\begin{align}
\Pi_{0} &\equiv	\sum_{s_1,s_2} \langle \frac{1}{2}s_1;\frac{1}{2}s_2|00 \rangle v(p_2,s_2)\bar{u}(p_1;s_1)=\frac{1}{4\sqrt{2N_{c}}E(E + m)}(\slashed{p}_2 - m)\gamma_5 
	\frac{\slashed{P} + 2E}{2E}(\slashed{p}_1 + m), \label{eq:gamma5_mod}
	\\
\Pi_{1} &\equiv	\sum_{s_1,s_2} \langle \frac{1}{2}s_1;\frac{1}{2}s_2|1S_z \rangle v(p_2,s_2)\bar{u}(p_1;s_1)=\frac{1}{4\sqrt{2N_{c}}E(E + m)}(\slashed{p}_2 - m)\gamma_{\mu}
	\frac{\slashed{P} + 2E}{2E}(\slashed{p}_1 + m), \label{eq:phi_sz_mod}
\end{align}
where $\Pi_{0}$ and $\Pi_{1}$ denote the projection operators for the spin‑singlet and spin‑triplet states, respectively, and Dirac spinors are normalized as $u^\dagger u=v^\dagger v=1$. \(N_{c} = 3\) is the number of colors.
We define \(v^{2} \equiv \frac{\mathbf{q}^2}{m^{2}}\) and expand the $S$-wave part of the amplitude \(\mathcal{M}\) in powers of \(v^{2}\)~\cite{Bodwin:2003wh}

\begin{equation}
\mathcal{M}_{S\text{-wave}} = \mathcal{M}_0 + \frac{\mathbf{q}^2}{m^2} \mathcal{M}_2 + \frac{\mathbf{q}^4}{m^4} \mathcal{M}_4 + O \left( \frac{\mathbf{q}^6}{m^6} \right), \label{MMEX}
\end{equation}
where
\begin{equation}
\begin{aligned}
&\mathcal{M}_0 = (\mathcal{M})_{\mathbf{q} \to 0}, \\
&\mathcal{M}_2 = \frac{m^2}{2!(d-1)} I^{\alpha \beta} \left( \frac{\partial^2 \mathcal{M}}{\partial q^\alpha \partial q^\beta} \right)_{\mathbf{q} \to 0}, \\
&\mathcal{M}_4 = \frac{m^4}{4!(d-1)(d+1)} I^{\alpha \beta \gamma \delta} \left( \frac{\partial^4 \mathcal{M}}{\partial q^\alpha \partial q^\beta \partial q^\gamma \partial q^\delta} \right)_{\mathbf{q} \to 0}.
\end{aligned} \label{M024}
\end{equation}
The tensor is defined as
\begin{equation}
	I^{\alpha \beta} = - g^{\alpha \beta} + \frac{P^{\alpha} P^{\beta}}{4 E^2}, \label{eq:I_alphabeta}
\end{equation}
\begin{equation}
	I^{\alpha \beta \gamma \delta} = I^{\alpha \beta} I^{\gamma \delta} + I^{\alpha \gamma} I^{\beta \delta} + I^{\alpha \delta} I^{\beta \gamma}. \label{eq:I_alphabeta_gamma_delta}
\end{equation}
In the axial gauge, the gluon propagator is defined as
\begin{equation}
	D^{\mu\nu}(l)
	=
	\frac{i}{l^{2}}
	\left(
	-g^{\mu\nu}
	+
	\frac{l^{\mu}\hat{n}^{\nu} + l^{\nu}\hat{n}^{\mu}}{l\!\cdot\! \hat{n}}
	\right),
\end{equation}
where the gluon momentum $ l=p_2+q_1$.

The fragmentation function for a heavy quark $Q$ fragmenting into a quark-antiquark pair state $(Q\bar{Q})$ is defined as
\begin{align}\label{eq:fragmentation_function}
	D_{{^1S_0}({^3S_1})/Q}(z)
	&=
	2E
	\int
	\frac{dq_{1}^{+}\, d^{2}\mathbf{q_{1T}}}
	{(2\pi)^{3}2q_{1}^{+}}
	\,
	2\pi\delta(k^{+}-P^{+}-q_{1}^{+})
	F_{c}\times\mathcal{A}_{Q \to (Q\bar{Q})[{}^{2S+1}L_{J}]}.
\end{align}
Here, the factor \(2E\) accounts for the relativistic normalization of the final-state hadron. Let \(k\), \(P\), and \(q_1\) denote the four-momentum of the fragmenting heavy quark, the four-momentum of the quark-antiquark pair \((Q\bar{Q})\), and the four-momentum of the final-state free quark, respectively. The color factor is given by \(F_{c} = \mathrm{Tr}(T^{a}T^{a}T^{b}T^{b}) = \frac{16}{3}\), and  \(\mathcal{A}_{Q \to (Q\bar{Q})[{}^{2S+1}L_{J}]}\) denotes the squared amplitude from the cut diagram for the fragmentation of a heavy quark \(Q\) into the state \((Q\bar{Q})[{}^{2S+1}L_{J}]\).
\begin{equation}
	\mathcal{A}_{Q \to (Q\bar{Q})[{}^{2S+1}L_{J}]}=
	\mathrm{Tr}\!\left[
	\slashed{\hat{n}}\,
	\mathcal{M}_{S-wave}(\slashed{q}_{1}+m)
	\mathcal{M}_{S-wave}^{*}
	\right].\label{eq:invariant_amplitude}
\end{equation}
\begin{align}
	\mathcal{M}_{S-wave}
	&=
	D_{\mu\nu}(l)	\frac{	\slashed{P}+\slashed{q}_{1}+m}{(P + q_{1})^{2}-m^{2}}
	\gamma^{\mu}
	\Pi_{0(1)}
	\gamma^{\nu},
	\\[6pt]
	\mathcal{M}_{S-wave}^{*}
	&=
	D^*_{\alpha\beta}(l)
	\gamma^{\alpha}
	\Pi_{0(1)}^{\dagger}
	\gamma^{\beta}
	\frac{	\slashed{P}+\slashed{q}_{1}+m}{(P + q_{1})^{2}-m^{2}}.
\end{align}
%The tensor  \(\Xi\)  is defined as
%\begin{equation}
%	\Xi=\left(
%	-g^{\mu\nu}
%	+
%	\frac{l^{\mu}\hat{n}^{\nu}+l^{\nu}\hat{n}^{\mu}}{l\!\cdot\!\hat{n}n}
%	\right)
%	\left(
%	-g^{\alpha\beta}
%	+
%	\frac{l^{\alpha}\hat{n}^{\beta}+l^{\beta}\hat{n}^{\alpha}}{l\!\cdot\! \hat{n}}
%	\right).
%\end{equation}
We obtain the LO result by setting $q = 0$ and $E = m$, and derive the relativistic corrections at $\mathcal{O}(v^2)$ and $\mathcal{O}(v^4)$ by expanding in $\mathbf{q}$. We perform the expansion at the squared amplitude level, with details provided in Appendix A. Our $\mathcal{O}(v^{2})$ results agree with Ref.~\cite{Sang:2009zz}, and the $\mathcal{O}(v^{4})$ corrections are new results presented in this work.

\section{ Matching the short-distance coefficients}
\label{Matching}

%We need to use Eq. \eqref{ex} to match the fragmentation function calculated in the previous section with the NRQCD matrix elements at the same order, thereby obtaining the SDCs.
By matching the results of perturbative NRQCD for the \(Q\bar{Q}\) production process with those of perturbative QCD obtained in the previous section, we derive the SDCs.
Under the standard normalization scheme, the NRQCD matrix elements are given as
\begin{equation}
	\langle \mathcal{O}^{Q\bar{Q}} (^1 S_0) \rangle = 2N_c,
\end{equation}
\begin{equation}
	\langle \mathcal{O}^{Q\bar{Q}} (^3 S_1) \rangle = 2(d-1)N_c,
\end{equation}
\begin{equation}
	\langle \mathcal{P}^{Q\bar{Q}} (n) \rangle = \boldsymbol{\mathbf{q}}^2 \langle \mathcal{O}^{Q\bar{Q}} (n) \rangle,
\end{equation}
\begin{equation}
	\langle \mathcal{Q}^{Q\bar{Q}} (n) \rangle = \boldsymbol{\mathbf{q}}^4 \langle \mathcal{O}^{Q\bar{Q}} (n) \rangle,
\end{equation}
where \( n = {}^1S_0, {}^3 S_1 \). The fragmentation function in the NRQCD factorization formula can be expressed as
\begin{equation}
D_{{ Q \bar{Q}} (^1 S_0) / Q}(z) = 2N_c \left( F^{^1S_0}(z) + \mathbf{q}^2 G^{^1S_0}(z) +\mathbf{q}^4 (H_1^{^1S_0}(z)+H_2^{^1S_0}(z)) \right),
\end{equation}
\begin{equation}
D_{{ Q \bar{Q}} (^3 S_1) / Q}(z) = 2N_c (d-1) \left( F^{^3 S_1}(z) + \mathbf{q}^2 G^{^3 S_1}(z) +\mathbf{q}^4 (H_1^{^3 S_1}(z) +H_2^{^3 S_1}(z) )\right).
\end{equation}
Through the above equation, we can determine the SDCs. For the ${}^1S_0$ state, we have
\begin{align}
&F^{^1S_0} = \frac{16\alpha_s^2 z(1 - z)^2 }{243m^3(2 - z)^6}(3 z^4-8 z^3+8 z^2+48), \label{F1S0}\\
&G^{^1S_0} =- \frac{8\alpha_s^2 z(1 - z)^2}{729m^5(2 - z)^8} (15 z^6-148 z^5+268 z^4-128 z^3+80 z^2-2496 z+2112),\label{G1S0} \\
&H_1^{^1S_0}+H_2^{^1S_0} = \frac{ 2 \alpha_s^2 z  (1 - z)^2}{54675 m^7 (2 - z)^{10}}\nonumber
\\
&\quad \quad\quad\quad\quad\quad\quad\times (5485 z^8-60352 z^7+272576 z^6-254176 z^5-473568 z^4
\nonumber\\
&\quad \quad\quad\quad\quad\quad\quad ~~~+500480 z^3+4723200 z^2-8240640 z+3528960).\label{eq:1S0}
\end{align}
 For the  ${}^3S_1$ state, we have
\begin{align}
&F^{^3S_1} = \frac{16\alpha_s^2 z(1 - z)^2 }{243m^3(2 - z)^6}(5 z^4-32 z^3+72 z^2-32 z+16), \label{F3S1} \\
&G^{^3S_1} =- \frac{8\alpha_s^2 z(1 - z)^2}{2187m^5(2 - z)^8} (183 z^6-2092 z^5+8924 z^4-18176 z^3+14416 z^2-5184 z+1344) , \label{G3S1}\\
&H_1^{^3S_1}+H_2^{^3S_1} = \frac{ 2 \alpha_s^2 z  (1 - z)^2}{164025 m^7 (2-z)^{10}}\nonumber \\
&\quad \quad\quad\quad\quad\quad\times (97885 z^8-1472296 z^7+9419408 z^6-32193088 z^5+62900576 z^4\nonumber\\
&\quad\quad\quad\quad\quad\quad~~~ -64604800 z^3+33583360 z^2-9369600 z+1639680).\label{eq:3S1}
\end{align}

Eqs.~\eqref{F1S0}, \eqref{G1S0}, \eqref{F3S1}, and \eqref{G3S1} agree with the results in Ref.~\cite{Sang:2009zz}, while our results in Eqs.~\eqref{eq:1S0} and \eqref{eq:3S1} are new.

\section{Numerical results and discussion}
\label{Num}
In this section, based on the NRQCD factorization framework, we perform a numerical analysis of the computed $\mathcal{O}(v^4)$ relativistic corrections to the $S$-wave fragmentation functions.
The relative values of the LDMEs in Eq.~\eqref{ex} are given by the following relations\cite{Bodwin:2014bia}, 
\begin{equation}
\langle 0| \mathcal{O}^{H} |0 \rangle : \langle 0| \mathcal{P}^{H} |0 \rangle : \langle 0| \mathcal{Q}_{1,2}^{H} |0 \rangle = 1 : m^2 \langle v^2 \rangle : m^4 \langle v^2 \rangle^2 
\end{equation}
The matrix elements $\langle v^2\rangle$ adhere to the velocity power scaling rules and are of the order of $v^2$ as follows,
\begin{equation}
	v^2=\langle v^2\rangle[1+\mathcal{O}(v^4)].
\end{equation}
The velocity parameter $v^2$ for the color-singlet(CS) state is estimated via the Gremm--Kapustin relation~\cite{Gremm:1997dq}:
\begin{equation}
	v^2 = \frac{M_{Q\bar{Q}} - 2m_{Q}^{\mathrm{pole}}}{m_{Q}^{\mathrm{QCD}}},
	\label{eq:v_squared}
\end{equation}
where \( m_{Q}^{\mathrm{pole}} \) is the pole mass of the quark, \( m_{Q}^{\mathrm{QCD}} \) is the quark mass in NRQCD theory, and \( M_{Q\bar{Q}} \) is the mass of the heavy quarkonium.
For a direct comparison between the $\mathcal{O}(v^4)$ and $\mathcal{O}(v^2)$ relativistic corrections, we employ the same values for the heavy quarkonium mass $M_{Q\bar{Q}}$ and the quark pole mass ${m_{Q}^{\mathrm{pole}}}$ as in Ref.~\cite{Sang:2009zz}, and take \( m_{Q}^{\mathrm{QCD}}=m_{Q}^{\mathrm{pole}} \). The values of meson mass are taken from Ref.~\cite{ParticleDataGroup:2008zun}.
\begin{equation}
m_{c}^{\mathrm{pole}} = 1.4~\mathrm{GeV}, \qquad
m_{b}^{\mathrm{pole}} = 4.6~\mathrm{GeV}.
\end{equation}
\begin{equation}
	\begin{aligned}
		&M_{\eta_c}       = 2.980~\text{GeV},          &&  ~~~~M_{\eta_b}       = 9.300~\text{GeV},\\
		&M_{J/\psi}       = 3.097~\text{GeV},          &&  ~~~~M_{\Upsilon(1S)} = 9.460~\text{GeV}.
	\end{aligned}
	\label{eq:masses}
\end{equation}
Therefore we obtain
\begin{align}
	\begin{aligned}
		&v_{\eta_c}^2 = 0.129, && v_{\eta_b}^2 = 0.022, \\
		&v_{J/\psi}^2 = 0.212, && v_{\Upsilon(1S)}^2 = 0.057.
	\end{aligned} 
	\label{eq:v_values}
\end{align}

With the parameters specified above, we present in this section the numerical distributions of the $S$-wave quarkonium fragmentation functions as a function of the momentum fraction $z$, and compare the $\mathcal{O}(v^{2})$ and $\mathcal{O}(v^{4})$ corrections with the LO result, as shown in Figure 1.

\begin{figure}[htbp]
	\centering
	
	\begin{subfigure}{0.48\textwidth}
		\centering
		\includegraphics[width=\linewidth]{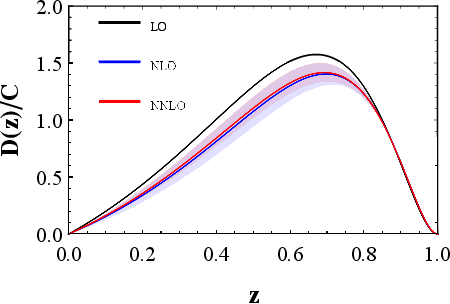}
		\caption{$c \to \eta_c$}
		\label{fig:eta_c_mc14_v2}
	\end{subfigure}
	\hfill
	\begin{subfigure}{0.48\textwidth}
	\centering
	\includegraphics[width=\linewidth]{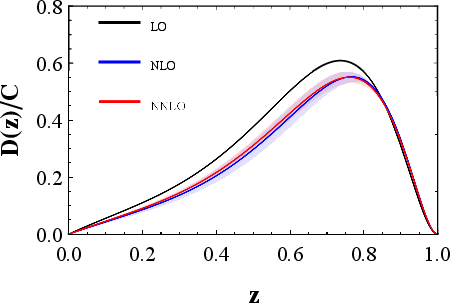}
	\caption{$c \to J/\psi$}
	\label{fig:jpsi_mc13_v2}
    \end{subfigure}
	
	\vspace{0.5cm}

		\begin{subfigure}{0.48\textwidth}
		\centering
		\includegraphics[width=\linewidth]{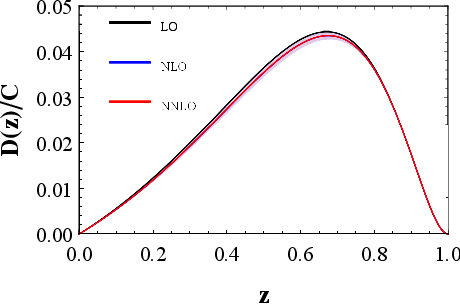}
		\caption{$b \to \eta_b$}
		\label{fig:eta_b_mb46_v2}
	\end{subfigure}
	\hfill
	\begin{subfigure}{0.48\textwidth}
		\centering
		\includegraphics[width=\linewidth]{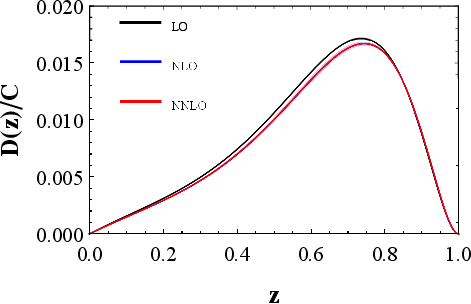}
		\caption{$b \to \Upsilon(1S)$}
		\label{fig:upsilon_mb48_v2}
	\end{subfigure}
	
	\caption{
		(Color online) The heavy-quark fragmentation functions $D(c \to \eta_c)$, $D(b \to \eta_b)$, $D(c \to J/\psi)$, and $D(b \to \Upsilon(1S))$ as functions of the momentum fraction $z$ in different $v^2$ ranges. The black, blue, and red curves correspond to the theoretical predictions at leading order (LO), next-to-leading order (NLO) $\mathcal{O}(v^2)$, and next-to-next-to-leading order (NNLO) $\mathcal{O}(v^4)$, respectively. The shaded bands accompanying each curve represent the theoretical uncertainties due to variations of $v^2$ within its adopted ranges. The calculations employ $v_{\eta_c}^2$, $v_{\eta_b}^2$, $v_{J/\psi}^2$, and $v_{\Upsilon(1S)}^2$ with heavy-quark masses fixed to $m_c = 1.4 \pm 0.05\,\text{GeV}$ and $m_b = 4.6 \pm 0.05\,\text{GeV}$, respectively. The normalization factor is taken as $C = 10^{-2}\, \alpha_s^2 \langle \mathcal{O} \rangle$.
	}
	\label{fig:fragmentation_functions}
\end{figure}

From the Figure 1, it can be observed that the $\mathcal{O}(v^{2})$ correction yields a significant negative contribution across most of the momentum fraction region, while the $\mathcal{O}(v^{4})$ correction provides a positive contribution, but with a magnitude that is notably smaller and generally much weaker than that of the $\mathcal{O}(v^{2})$ term. The results clearly demonstrate a good convergence behavior of the relativistic expansion. This convergence is consistent with the findings in Refs.~\cite{Bodwin:2012xc} for the relativistic corrections in gluon fragmentation into ${}^{3}S_{1}$ states, where the $\mathcal{O}(v^{4})$ contribution was also observed to be much smaller than the $\mathcal{O}(v^{2})$ contribution, and an infrared structure was first identified at $\mathcal{O}(v^{4})$ in gluon fragmentation into ${}^{3}S_{1}$. Our numerical conclusion for the fragmentation functions of heavy quarks into ${}^{1}S_{0}$ and ${}^{3}S_{1}$ states follows this trend, though no divergence was found in our calculation.

To further quantify the contributions of different orders of relativistic corrections relative to LO, we present the ratio $\mathcal{O}(v^{4})/\mathrm{LO}$ in Table~\ref{table}. 
Additionally, we provide in Table~\ref{table} the ratios of relativistic correction SDCs to the LO SDCs in the limits $z \to 0$ and $z \to 1$.The limit $z \to 0$ corresponds to the small-$z$ region, where the quarkonium carries only a small fraction
of the parent parton momentum and probes the near-threshold (soft) production dynamics.
In contrast, the limit $z \to 1$ represents the forward (collinear) fragmentation region, in which the quarkonium
inherits almost the entire momentum of the fragmenting parton.
It can be seen that the $\mathcal{O}(v^{4})$ correction generally makes a small contribution relative to LO, amounting to only a few percent or less in most $z$ regions. Therefore, in high-precision predictions of $S$-wave heavy quarkonium production in the large transverse momentum region, the $\mathcal{O}(v^{2})$ correction remains the dominant relativistic effect, while higher order corrections present a small, manageable contribution. The results indicate that higher-order relativistic corrections do not compromise the validity of the NRQCD expansion, and the relativistic expansion exhibits a good hierarchical structure in the $S$-wave system.

\begin{table}[H]
	\centering
	\caption{Integrated ratios of relativistic corrections. Here, $D^{(0)}(z)$ denotes the LO fragmentation function, $D^{(2)}(z)$ the NLO relativistic correction term, and $D^{(4)}(z)$ the NNLO relativistic correction term. The ratios in the table quantify the overall magnitude of the relativistic corrections relative to the LO contribution. The up and down labels for each result represent the theoretical uncertainties due to variations of $v^2$ within its adopted ranges.}
	\label{table}
		\renewcommand{\arraystretch}{2}  % 增加行高，默认是1.0
\begin{tabular}{l c c c c} 
	\hline \hline 
	 & \(c \to \eta_{c}\) & \(c \to J/\psi\) & \(b \to \eta_{b}\) &\(b \to \Upsilon(1S)\) \\ 
	\hline 
	$ \frac{\int dzD^{(2)}(z)}{\int dzD^{(0)}(z)}$ & $-13.05^{+7.45}_{-8.00}\%$ & $-13.19^{+4.74}_{-5.09}\%$ & $-2.21^{+2.16}_{-2.25}\%$ & $-3.51^{+1.37}_{-1.41}\%$ \\ 
		\hline 
	$ \frac{\int dzD^{(4)}(z)}{\int dzD^{(2)}(z)}$ & \(-12.39^{+7.07}_{-7.60}\%\) & \(-12.83^{+4.61}_{-4.95}\%\) & \(-2.09^{+2.05}_{-2.14}\%\) & \(-3.42^{+1.34}_{-1.37}\%\) \\ 
	\hline 
	$ \frac{\int dzD^{(4)}(z)}{\int dzD^{(0)}(z)}$ & \(1.62^{+2.59}_{-1.32}\%\) & \(1.69^{+1.56}_{-1.00}\%\) & \(0.05^{+0.14}_{-0.05}\%\) & \(0.12^{+0.12}_{-0.08}\%\) \\ 

	\hline \hline 
\end{tabular}
\end{table}

\begin{table}[htbp]
	\centering
\caption{
	Ratios of relativistic correction SDCs to the LO SDCs in the limits $z \to 0$ and $z \to 1$. Here $R_{1,2}$ and $R_{3,4}$ correspond to $\mathcal{O}(v^2)$ and $\mathcal{O}(v^4)$ corrections, respectively, with $R_{1,2} = \left.\frac{m^2G(n)}{F(n)}\right|_{z \to 0,1}$ and
	$R_{3,4} = \left.\frac{m^4[H_1(n)+H_2(n)]}{F(n)}\right|_{z \to 0,1}$.
}
	\label{tab:relativistic_corrections}
	\renewcommand{\arraystretch}{1.5}  % 增加行高，默认是1.0
	\begin{tabular}{@{\extracolsep{1em}} l c c c c @{}}  % 添加列间距
		\hline
			\hline
		 & 
		$R_1$ & $R_2$ & $R_3$ & $R_4$ \\
		& $(z \to 0)$ & $(z \to 1)$ & $(z \to 0)$ & $(z \to 1)$ \\ 
		\hline
		$^1S_0$ & $-1.83$  &  $0.97$ &  $2.55$ &  $0.02$ \\ 
			\hline
				$^3S_1$ & $-1.17$  &  $1.12$ &  $1.19$ &  $0.01$ \\ 
				\hline
				\hline
	\end{tabular}
\end{table}

%\newpage
\section{Summary}
\label{summary}
Based on the NRQCD factorization framework and the gauge-invariant definition of the fragmentation function introduced by Collins--Soper, we have systematically computed the relativistic expansion of the quark fragmentation process into $S$-wave heavy quarkonium states. Building upon existing calculations of the $\mathcal{O}(v^{2})$ relativistic corrections, we have further completed the computation of the $\mathcal{O}(v^{4})$ SDCs for the $S$-wave quarkonium fragmentation functions.
In Appendix B, we have further extended the calculation to the unequal-mass case and obtained the analytical expressions for the $\mathcal{O}(v^{4})$ SDCs. 
 Our results demonstrate that the relativistic expansion of heavy quark fragmentation into ${}^{1}S_{0}$ and ${}^{3}S_{1}$ states exhibits good convergence. The $\mathcal{O}(v^{2})$ term is essential for improving predictive accuracy, while the $\mathcal{O}(v^{4})$ term, though small, remains meaningful for high-precision descriptions. The $\mathcal{O}(v^{4})$ SDCs provided in this work fill a gap in the literature and can be utilized for future high-precision predictions of $\eta_c$ and $J/\psi$ production rates in the large transverse momentum region.

\section{Acknowledgements} 
This work was supported by the National Natural Science Foundation of China (No. 11705078, 12575087).

\appendix
\section*{Appendix}
\label{app:Bc_fragmentation}
\setcounter{equation}{0}  % 重置公式计数器
\renewcommand{\theequation}{A.\arabic{equation}}  % 公式编号格式为 A.1, A.2...
\subsection*{A. Order-by-order expansion of the squared amplitude}
\label{app:amplitude}
In Appendix A, we present the method for computing the order‑by‑order expansion of the squared amplitude.
For convenience of performing the $v$ expansion ,
we rewrite the amplitude corresponding to the one in Eq.~\eqref{eq:invariant_amplitude} 
\begin{equation}
	\mathcal{M}(q) \equiv \widetilde{\mathcal{M}}\!\left(q,\, E(\mathbf{q})\right),
\end{equation}
where $\widetilde{\mathcal{M}}(q,E)$ is regarded as a function of the two independent variables $q$ and $E$.
In the limit $q \rightarrow 0$, one has $E(\mathbf{q}) \rightarrow m$, and therefore
\begin{equation}
	\mathcal{M}_0 \equiv \widetilde{\mathcal{M}}(0,\, m).
\end{equation}
The relativistic dispersion relation is expanded as
\begin{equation}
	E(\mathbf{q}) = m + \frac{\mathbf{q}^2}{2m} - \frac{\mathbf{q}^4}{8m^3} + O\!\left(\frac{\mathbf{q}^6}{m^5}\right),
\end{equation}
Following Eq.~\eqref{MMEX}, the $S$-wave component of the squared amplitude, in which trace evaluations and other intermediate algebraic steps are omitted, is expanded in powers of $v$ as
\begin{align}
	|\mathcal{M}|^2 
	&= \mathcal{M}_0 \mathcal{M}_0^*
	+ \frac{\mathbf{q}^2}{m^2}\bigl(\mathcal{M}_0 \mathcal{M}_2^* + \text{h.c.}\bigr)
	+ \frac{\mathbf{q}^4}{m^4}\Bigl[\mathcal{M}_2 \mathcal{M}_2^* + \bigl(\mathcal{M}_0 \mathcal{M}_4^* + \text{h.c.}\bigr)\Bigr]
	+ O(v^6)                                                                  \nonumber \\[6pt]
&= \widetilde{\mathcal{M}}(0,m)\,\widetilde{\mathcal{M}}^*(0,m)                                           \nonumber \\[6pt]
&\quad
+ \frac{\mathbf{q}^2}{6}\, I^{\alpha\beta}
\Bigl[
\frac{\partial^2 \widetilde{\mathcal{M}}}{\partial q^\alpha \partial q^\beta}(0,m)\,\widetilde{\mathcal{M}}^*(0,m)
+ \text{h.c.}
\Bigr]                                                                   
+ \frac{\mathbf{q}^2}{2m}\,
\frac{\partial\!\left(\widetilde{\mathcal{M}}(0,m)\,\widetilde{\mathcal{M}}^*(0,m)\right)}{\partial E}              \nonumber \\[6pt]
&\quad
+ \frac{\mathbf{q}^4}{36}\,
I^{\alpha\beta} I^{\gamma\delta}
\Bigl[
\frac{\partial^2 \widetilde{\mathcal{M}}}{\partial q^\alpha \partial q^\beta}(0,m)\,
\frac{\partial^2 \widetilde{\mathcal{M}}^{*}}{\partial q^\gamma \partial q^\delta}(0,m)
\Bigr]                                                                     \nonumber \\[6pt]
&\quad
+ \frac{\mathbf{q}^4}{360}\,
I^{\alpha\beta\gamma\delta}
\Bigl[
\frac{\partial^4 \widetilde{\mathcal{M}}}{\partial q^\alpha \partial q^\beta \partial q^\gamma \partial q^\delta}(0,m)\,
\widetilde{\mathcal{M}}^*(0,m)
+ \text{h.c.}
\Bigr]                                                                      \\[6pt]
&\quad
+ \frac{\mathbf{q}^4}{12m}\,
\frac{\partial}{\partial E}
\Bigl[
I^{\alpha\beta}
\frac{\partial^2 \widetilde{\mathcal{M}}}{\partial q^\alpha \partial q^\beta}(0,E)\,\widetilde{\mathcal{M}}^*(0,m)
+ \text{h.c.}
\Bigr]                                                                     \nonumber \\[6pt]
&\quad
+ \frac{\mathbf{q}^4}{8m^2}\,
\frac{\partial^2\!\left(\widetilde{\mathcal{M}}(0,m)\,\widetilde{\mathcal{M}}^*(0,m)\right)}{\partial E^2}         
- \frac{\mathbf{q}^4}{8m^3}\,
\frac{\partial\!\left(\widetilde{\mathcal{M}}(0,m)\,\widetilde{\mathcal{M}}^*(0,m)\right)}{\partial E}              \nonumber \\[6pt]
&\quad
+ O(v^6).                                                                 \nonumber
\end{align}

\setcounter{equation}{0}  % 重置公式计数器
\renewcommand{\theequation}{B.\arabic{equation}}  % 公式编号格式为 A.1, A.2...

\subsection*{B.$\mathcal{O}(v^{4})$ relativistic corrections for heavy‑quark fragmentation into unequal‑mass $S$-wave mesons }
\label{app:unequal}
In Appendix B, we extend the computation of $\mathcal{O}(v^{4})$ relativistic corrections for heavy‑quark fragmentation into equal‑mass $S$‑wave quarkonia to systems with unequal masses, exemplified by the $B_c$ meson. For heavy flavored mesons, the spin‑singlet and spin‑triplet projection operators read
\begin{align}
	\Pi_{0} &=	\frac{1}{4\sqrt{2N_{c}}\sqrt{E_1E_2(E_1 + m_1)(E_2 + m_2)}}(\slashed{p}_2 - m_2)\gamma_5 
	\frac{\slashed{P} + E_1+E_2}{E_1+E_2}(\slashed{p}_1 + m_1), \label{eq:gamma5_mod}
	\\
	\Pi_{1} &=	\frac{1}{4\sqrt{2N_{c}}\sqrt{E_1E_2(E_1 + m_1)(E_2 + m_2)}}(\slashed{p}_2 - m_2)\gamma_\mu 
	\frac{\slashed{P} + E_1+E_2}{E_1+E_2}(\slashed{p}_1 + m_1), \label{eq:phi_sz_mod}
\end{align}
where \(m_1\) and \(m_2\) are the masses of the two quarks in the meson. The momenta of the two quarks, \(p_1\) and \(p_2\), can be expressed in terms of the meson four-momentum \(P\) and the half‑relative momentum \(q\) of the two quarks inside the meson as
\begin{align}
	p_1 &= rP + q, \\
	p_2 &= (1 - r)P - q, 
\end{align}
where
\begin{align}
		r &= \frac{E_1}{E_1 + E_2},\\
	E_1 &= \sqrt{\mathbf q^2 + m_1^2}, \\
	E_2 &= \sqrt{\mathbf q^2 + m_2^2}.
\end{align}
Similar to the equal‑mass case, we expand \(E_1\) and \(E_2\) in powers of \(\mathbf q^2\).
We obtain the $\mathcal{O}(v^4)$ SDCs for the unequal‑mass case. For the ${}^1S_0$ state, we have
\begin{align}
	&F^{^1S_0} =\frac{4\alpha_s^2z(1 - z)^2}{243M^3(y_1 - 1)^2(y_1z - 1)^6} \nonumber\\
	&\quad~~~~\times \left[\left(6y_1^4-6y_1^3+3y_1^2\right) z^4+\left(-36y_1^3+34y_1^2-10y_1\right) z^3+\left(68y_1^2-62y_1+15\right) z^2+(18-36y_1) z+6\right],\label{eq:F1S0} \\
	&G^{^1S_0} =-\frac{\alpha_s^2z(1 - z)^2}{729M^5(y_1 - 1)^4 y_1^2(y_1z - 1)^8} \nonumber \\
	&\quad~~~~\times\big[\left(108y_1^8-264y_1^7+372y_1^6-240y_1^5+57y_1^4\right) z^6\nonumber \\
	&\quad~~~~ +\left(-960y_1^7+2204y_1^6-2588y_1^5+1466y_1^4-320y_1^3\right) z^5\nonumber \\
	&\quad~~~~ +\left(3004y_1^6-6716y_1^5+7186y_1^4-3636y_1^3+690y_1^2\right) z^4\nonumber \\
	&\quad~~~~ +\left(-3776y_1^5+7808y_1^4-8000y_1^3+3840y_1^2-664y_1\right) z^3\nonumber \\
	&\quad~~~~ +\left(1780y_1^4-2456y_1^3+2040y_1^2-716y_1+45\right) z^2\nonumber \\
	&\quad~~~~ +\left(-384y_1^3+180y_1^2-180y_1+54\right) z\nonumber \\
	&\quad~~~~ +36y_1^2+12y_1+18  \big], \label{eq:G1S0}\\
	&H_1^{^1S_0}+H_2^{^1S_0} = \frac{  \alpha_s^2 z  (1 - z)^2}{218700 M^7 (y_1-1)^6 y_1^4 (y_1 z-1)^{10}}\nonumber
	\\
	&\quad~~~~\times \big[\left(40500y_1^{12}-154620y_1^{11}+353230y_1^{10}-454280y_1^9+335200y_1^8-132990y_1^7+21945y_1^6\right) z^8 \nonumber \\
	&\quad~~~~+\left(-471240y_1^{11}+1693980y_1^{10}-3398020y_1^9+3980908y_1^8-2755624y_1^7+1043814y_1^6-165698y_1^5\right) z^7 \nonumber \\
	&\quad~~~~+\left(2062880y_1^{10}-7225900y_1^9+13489206y_1^8-14700848y_1^7+9534738y_1^6-3416386y_1^5+516875y_1^4\right) z^6 \nonumber \\
	&\quad~~~~+\left(-4185560y_1^9+14202964y_1^8-25790632y_1^7+27208892y_1^6-16960684y_1^5+5794810y_1^4-826860y_1^3\right) z^5 \nonumber \\
	&\quad~~~~+\left(4063752y_1^8-12419196y_1^7+22131686y_1^6-23224112y_1^5+14365390y_1^4-4809970y_1^3+656175y_1^2\right) z^4 \nonumber \\
	&\quad~~~~+\left(-1845560y_1^7+3846180y_1^6-6342180y_1^5+6470340y_1^4-3869840y_1^3+1187050y_1^2-128850y_1\right) z^3 \nonumber \\
	&\quad~~~~+\left(500480y_1^6-457140y_1^5+927530y_1^4-1094640y_1^3+680750y_1^2-180750y_1+10125\right) z^2 \nonumber \\
	&\quad~~~~+\left(-89640y_1^5+8940y_1^4-127440y_1^3+148500y_1^2-78300y_1+12150\right) z \nonumber \\
	&\quad~~~~+8100y_1^4+2520y_1^3+12300y_1^2-9000y_1+4050 \big]\label{eq:H1S0}.
\end{align}

For the  ${}^3S_1$ state, we have
\begin{align}
	&F^{^3S_1} = \frac{4 \alpha_s^2 z  (1 - z)^2}{243M^3(y_1-1)^2(y_1z-1)^6} \nonumber\\
	&\quad~~~~\times \left[\left(2y_1^4-2y_1^3+3y_1^2\right) z^4+\left(-4y_1^3+6y_1^2-10y_1\right) z^3+\left(12y_1^2-18y_1+15\right) z^2+(-4y_1-2) z+2\right],\label{eq:F3S1} \\
	&G^{^3S_1}=-\frac{ \alpha_s^2 z  (1 - z)^2}{2187M^5(y_1 - 1)^4 y_1^2 (y_1 z - 1)^8} \nonumber \\
	&\quad~~~~\times\big[\left(156y_1^8-360y_1^7+624y_1^6-492y_1^5+171y_1^4\right) z^6\nonumber
	\\
	&\quad~~~~+\left(-528y_1^7+1532y_1^6-3172y_1^5+2666y_1^4-960y_1^3\right) z^5\nonumber
	\\
	&\quad~~~~+\left(1516y_1^6-4476y_1^5+7818y_1^4-6004y_1^3+2070y_1^2\right) z^4\nonumber
	\\
	&\quad~~~~+\left(-2144y_1^5+5328y_1^4-8064y_1^3+5816y_1^2-1992y_1\right) z^3\nonumber
	\\
	&\quad~~~~+\left(868y_1^4-104y_1^3-476y_1^2+336y_1+135\right) z^2\nonumber
	\\
	&\quad~~~~+\left(-528y_1^3+420y_1^2-204y_1-18\right) z\nonumber
	\\
	&\quad~~~~+84y_1^2-36y_1+18\big], \label{eq:G3S1}\\
	&H_1^{^3S_1}+H_2^{^3S_1} = \frac{  \alpha_s^2 z  (1 - z)^2}{656100 M^7 (y_1-1)^6 y_1^4 (y_1 z-1)^{10}}\nonumber
	\\
	&\quad~~~~\times \big[\left(72420y_1^{12}-259500y_1^{11}+612490y_1^{10}-842000y_1^9+698500y_1^8-320790y_1^7+65835y_1^6\right) z^8\nonumber
	\\
	&\quad~~~~+\left(-350040y_1^{11}+1470220y_1^{10}-4009740y_1^9+5920524y_1^8-5111752y_1^7+2398182y_1^6-497094y_1^5\right) z^7\nonumber
	\\
	&\quad~~~~+\left(1129120y_1^{10}-4937100y_1^9+12964218y_1^8-18635224y_1^7+15907614y_1^6-7449138y_1^5+1550625y_1^4\right) z^6\nonumber
	\\
	&\quad~~~~+\left(-2399560y_1^9+9672692y_1^8-22918376y_1^7+31003676y_1^6-25673572y_1^5+11879250y_1^4-2480580y_1^3\right) z^5\nonumber
	\\
	&\quad~~~~+\left(2465256y_1^8-7682668y_1^7+16212858y_1^6-21281256y_1^5+18198730y_1^4-8848170y_1^3+1968525y_1^2\right) z^4\nonumber
	\\
	&\quad~~~~+\left(-1297960y_1^7+1269780y_1^6-190860y_1^5-653660y_1^4-401280y_1^3+869850y_1^2-386550y_1\right) z^3\nonumber
	\\
	&\quad~~~~+\left(833920y_1^6-1308020y_1^5+1376230y_1^4-674280y_1^3+178050y_1^2-31950y_1+30375\right) z^2\nonumber
	\\
	&\quad~~~~+\left(-227640y_1^5+291180y_1^4-307680y_1^3+155700y_1^2-33300y_1-4050\right) z\nonumber
	\\
	&\quad~~~~+25620y_1^4-23400y_1^3+27900y_1^2-16200y_1+4050\big]\label{eq:H3S1}.
\end{align}
where \( M = m_1 + m_2 \) and \( y_1 = \dfrac{m_1}{m_1 + m_2} \). Eqs.~\eqref{eq:F1S0}, \eqref{eq:G1S0}, \eqref{eq:F3S1}, and \eqref{eq:G3S1} are consistent with Ref.~\cite{Sang:2009zz} and Ref.~\cite{Yang:2019gga}, while Eqs.~\eqref{eq:H1S0} and \eqref{eq:H3S1} present new results.
Under the conditions \(y_1 = \dfrac{1}{2}\) and \(M = 2m\), the unequal‑mass \(\mathcal{O}(v^{4})\)  short‑distance coefficients given in Eqs.~\eqref{eq:H1S0} and~\eqref{eq:H3S1} reduce to the equal‑mass SDCs in Eqs.~\eqref{eq:1S0} and~\eqref{eq:3S1}, respectively. Taking the \( B_c \) meson as an example, we present the integrated ratios of the relativistic corrections in Table~\ref{tab:integrated_ratios}.
Notably, with $v^2 = 0.13$, the ratios of the integrated $\mathcal{O}(v^4)$ relativistic corrections to the LO contributions are $8.48\%$ for $b \to B_{c}^-(b\bar{c})$ and $7.56\%$ for $b \to B_{c}^{*-}(b\bar{c})$, respectively. In contrast, the corresponding ratios for the $\mathcal{O}(v^2)$ corrections are $-29.86\%$ and $-27.56\%$.

\begin{table}[H]
	\centering
	\caption{Integrated ratios of relativistic corrections for the \( B_c \) meson. }
	\label{tab:integrated_ratios}
	\renewcommand{\arraystretch}{2}  % 增加行高，默认是1.0
\begin{tabular}{@{\extracolsep{1em}} l c c c c @{}}
	\hline
	\hline
	& \multicolumn{2}{c}{$m_1=1.4,\;m_2=4.6$} & \multicolumn{2}{c}{$m_1=4.6,\;m_2=1.4$} \\
	\cline{2-5}
	& \(c \to B_{c}^+(c\bar{b})\) & \(c \to B_{c}^{*+}(c\bar{b})\) & \(b \to B_{c}^-(b\bar{c})\) & \(b \to B_{c}^{*-}(b\bar{c})\) \\
	\hline
	$ \frac{\int dzD^{(2)}(z)}{\langle v^2 \rangle\int dzD^{(0)}(z)}$ & $ -0.62$ & $-0.31$ & $ -2.30$ & $-2.12$ \\ 
	\hline
	$ \frac{\int dzD^{(4)}(z)}{\langle v^2 \rangle\int dzD^{(2)}(z)}$ & $ -1.53$ & $-1.83$ & $ -2.18$ & $-2.11$ \\ 
	\hline
	$ \frac{\int dzD^{(4)}(z)}{\langle v^4 \rangle\int dzD^{(0)}(z)}$  & $ 0.95$ & $0.57$ & $ 5.01$ & $4.47$ \\ 
	\hline
	\hline
\end{tabular}
\end{table}

%\newpage
% ========================
% 参考文献：完整 DOI 版
% ========================

\end{document}